\documentclass[aps,pra,twocolumn,showpacs,preprintnumbers,amsmath,amssymb]{revtex4}
\usepackage{graphicx}
\usepackage{bm}
\usepackage{dcolumn}

\def\sc#1#2#3{Science {\bf #1}, #2 (#3)}
\def\prv#1#2#3{Phys. Rev. {\bf #1}, #2 (#3)}
\def\rmp#1#2#3{Rev. Mod. Phys. {\bf #1}, #2 (#3)}
\def\prl#1#2#3{Phys. Rev. Lett. {\bf #1}, #2 (#3)}
\def\pra#1#2#3{Phys. Rev. A {\bf #1}, #2 (#3)}

\def\ajp#1#2#3{Am. J. Phys. {\bf #1}, #2 (#3)}

\def\noi{\noindent}
\def\bc{\begin{center}}
\def\ec{\end{center}}
\newcommand{\bea}{\begin{equation}}
\newcommand{\eea}{\end{equation}\noi}
\newcommand{\ber}{\begin{eqnarray}}
\newcommand{\eer}{\end{eqnarray}\noi}
\begin{document}
\title{Thermodynamics of quantum gases for the entire range of temperature}
\author{Shyamal Biswas}\email{sbiswas.phys.cu@gmail.com}
\author{Debnarayan Jana}
\affiliation{Department of Physics, University of Calcutta, 92 APC Road, Kolkata-700009, India}

\date{\today}

\begin{abstract}
We have analytically explored thermodynamics of free Bose and Fermi gases for the entire range of temperature, and have extended the same for harmonically trapped cases. We have obtained approximate chemical potentials of the quantum gases in closed forms of temperature so that the thermodynamic properties of the quantum gases become plausible specially in the intermediate regime between the classical and quantum limits.
\end{abstract}
\pacs{01.40.-d, 03.75.Hh, 03.75.Ss, 05.30.-d}

\maketitle
\section{Introduction}
In the undergraduate courses on statistical physics one has to plot specific heats of free (ideal homogeneous) Bose and Fermi gases \cite{pathria}. To plot specific heat of a quantum gas with respect to its temperature, one has to know temperature dependence of its chemical potential which in general, is obtained from an implicit relationship with the number density of particles \cite{pathria,pitaevskii-book}. This relationship involves a polylog function of an exponential of the chemical potential which can not be obtained in a closed form of temperature as because inverse of a polylog function, in general, does not exist. In spite of that, students are taught how to get approximate temperature dependence of the chemical potential for the classical and quantum (degenerate) regimes \cite{pathria}. But, for the intermediate regime, they are advised to get graphical solutions. Feeling difficulties of obtaining the graphical solutions, we in this paper, will incorporate an easy technique of obtaining approximate chemical potentials of the quantum gases in closed forms of temperature not only for the intermediate regime but also for the entire range of temperature.

After the observation of Bose-Einstein condensation in 1995 \cite{jila,rice,mit}, studying trapped quantum gases now-a-days becomes a topic of high experimental and theoretical interest of research \cite{pitaevskii-rmp1,bloch-rmp,pitaevskii-rmp2}. So, it is reasonable to extend our discussion for harmonically trapped (ideal inhomogeneous) cases.

Calculations of this article will begin with the Bose-Einstein statistics. In subsection-2A we will present a technique of how to get an approximate chemical potential of a free Bose gas in a closed form of temperature. We will do the same for a harmonically trapped Bose gas in subsection-2B, and for free and harmonically trapped Fermi gases in section-3. We will plot approximate chemical potentials in FIG 1. We will also plot specific heats, equation of states, and susceptibilities in FIGs 2 and 3 using our approximate chemical potentials.

\section{Thermodynamics of Bose gases}
Let us consider a 3-D free Bose gas in thermodynamic equilibrium with a (heat and particle) reservoir which is characterized by temperature $T$ and chemical potential $\mu$. All the thermodynamic quantities of this system can be obtained from the Bose-Einstein statistics
\begin{eqnarray}
\bar{n}_i=\frac{1}{\text{e}^{(\epsilon_i-\mu)/kT}-1},
\end{eqnarray}
where $\bar{n}_i$ is the (equilibrium) average number of particles at the $i$th single particle state having energy $\epsilon_i$. From Eqn.(1) we can see that total average number of particles ($N=\sum_i\bar{n}_i$) and total average energy ($E=\sum_i\bar{n}_i\epsilon_i$) are functions of $\mu$. Hence, all the thermodynamic quantities like pressure ($p$), average number density of particles ($\bar{n}=N/V$), average energy per particle ($\bar{\epsilon}$), specific heat per particle ($c_v$), etc. are expected to be functions of $\mu$.

For a free gas, we can replace the single particle energy eigenstates $\{i\}$ by the single particle momenta $\{\textbf{p}\}$ so that $\epsilon_i$ essentially becomes $\epsilon_\textbf{p}=p^2/2m$ where $m$ is the mass of a single particle. Degeneracy of the level $\epsilon_\textbf{p}$ in the semiclassical limit is given by $\frac{V4\pi p^2dp}{(2\pi\hbar)^3}$, where $V$ is the volume of the system. In the thermodynamic limit ($N\rightarrow\infty,V\rightarrow\infty, N/V=const.$), standard text book result for specific heat per particle of the free Bose gas is given by \cite{pathria}
\bea
c_v=\bigg\{\begin{matrix}&k\big[\frac{15}{4}\frac{\text{Li}_{5/2}(z)}{\text{Li}_{3/2}(z)}-\frac{9}{4}\frac{\text{Li}_{3/2}(z)}{\text{Li}_{1/2}(z)}\big] \ \ \ \ \ \ \ \ \ \text{for} \ \ T>T_c\\
&k\big[\frac{15}{4}\frac{\zeta(5/2)}{\zeta(3/2)}\big]\big(\frac{T}{T_c}\big)^{3/2} \ \ \ \ \ \ \ \ \ \ \ \ \ \ \ \ \text{for} \ \ T\le T_c
\end{matrix}
\eea
where $T_c=\frac{2\pi\hbar^2}{mk}\big(\frac{\bar{n}}{\zeta(3/2)}\big)^{2/3}$ is the Bose-Einstein condensation temperature at and below which $\mu$ takes the highest possible value ($0$), $z=\text{e}^{\mu/kT}$ is the fugacity, and $\text{Li}_j(z)=z+\frac{z^2}{2^j}+\frac{z^3}{3^j}+...$ is a polylog function of order $j$. Plotting $c_v$ for $T\le T_c$ is very easy. But, plotting $c_v$ for $T>T_c$ is not an easy job until one manages to get the temperature dependence of $z$ or of $\mu$ from the implicit relation \cite{pathria}
\begin{eqnarray}
\bigg(\frac{\text{Li}_{3/2}(z)}{\zeta(3/2)}\bigg)^{2/3}=\frac{1}{t},
\end{eqnarray}
which can be obtained as a result of phase-space integration of the right side of Eqn.(1), and where $t=T/T_c$. As inverse of a polylog function ($\text{Li}_{j}(z)$) does not exist except for $j=1$, we are not able to get $\mu$ from Eqn.(3) or $c_v$ from Eqn.(2) as a function of temperature without approximation in particular for $T>T_c$.

On the other hand, for a harmonically trapped case, all the particles are 3-D harmonic oscillators, and the single particle energy levels are given by $\epsilon_i=(\frac{3}{2}+i)\hbar\omega$, where $\omega$ is the angular frequency of oscillations. Although the degeneracy ($g_i$) of this level is $i^2/2+3i/2+1$ \cite{powell}, yet in the  thermodynamic limit ($N\rightarrow\infty$, $\omega\rightarrow0$ \& $N\omega^3=const.$), only the first term of the degeneracy contributes significantly. Zero point energy can also be neglected in this limit. The implicit relation between the chemical potential and temperature (in the thermodynamic limit) for this case is given instead of Eqn.(3) by \cite{pitaevskii-book,pethick-book,pitaevskii-rmp1}
\begin{eqnarray}
\bigg(\frac{\text{Li}_{3}(z)}{\zeta(3)}\bigg)^{1/3}=\frac{1}{t},
\end{eqnarray}
where $t=\frac{T}{T_c}$ and $T_c=\frac{\hbar\omega}{k}[\frac{N}{\zeta(3)}]^{1/3}$ \cite{pitaevskii-book,pethick-book,pitaevskii-rmp1}.

In the following we will present approximate analyses of Eqns.(3) and (4), and will obtain approximate chemical potentials of the two cases indicating suitable suffices for different temperature regimes.

\subsection{For free Bose gas}
Classical regime of a quantum gas is characterized by $t\gg1$ (or by $z\ll1$) so that in this regime we may write $\text{Li}_{3/2}(z)\approx z$ which along with Eqn.(3) yields the approximate classical result
\bea
\mu_{\gg}(t)\approx-kT\frac{3}{2}\text{ln}\big(t/\zeta^{\frac{2}{3}}(3/2)\big).
\eea

On the other hand, the quantum regime for Bose gas, is characterized by $t\lesssim1$ (or by $z\sim1$). For $t\le1$, we have $\mu=0$ \cite{pitaevskii-book}, and for $t>1$, $\mu$ is to be obtained from Eqn.(3). In the following we will extend a standard (but not much familiar) result \cite{bhattacharjee} for $t\gtrapprox1$ up to the second lowest order. From Eqns.(1) and (3) we can redefine the polylog function $\text{Li}_{3/2}(z)$ as
\bea
\text{Li}_{3/2}(z)=\frac{2}{\sqrt{\pi}}\int_0^\infty\bigg[\frac{1}{\text{e}^{\epsilon+\nu}-1}\bigg]\epsilon^{1/2}\text{d}\epsilon,
\eea
in an integral form where $\nu=|\mu|/kT$. One can check that $\text{Li}_{3/2}(z)$ is not an analytic function of $z$ particularly about $z=1$. Thus to know even the approximate values of chemical potential for $t\gtrapprox1$ is a tricky job. Let us adopt the trick of Landau and Lifshitz for this purpose \cite{landau}. Subtracting $\text{Li}_{3/2}(1)$ from $\text{Li}_{3/2}(z)$ we can recast the above equation as \cite{landau}
\begin{eqnarray}
\text{Li}_{\frac{3}{2}}(z)-\zeta\big(\frac{3}{2}\big)=\frac{2}{\sqrt{\pi}}\int_0^\infty\bigg[\frac{1}{\text{e}^{\epsilon+\nu}-1}-\frac{1}{\text{e}^{\epsilon}-1}\bigg]\epsilon^{\frac{1}{2}}\text{d}\epsilon.
\end{eqnarray}
It is clear from Eqn.(7) that, for $t\gtrapprox1$ (or for $\nu\ll1$), the integrand will contribute only for smaller values of $\epsilon$, so that up to the second lowest order we can have
\begin{eqnarray}
\text{Li}_{\frac{3}{2}}(z)&\approx&\zeta\big(\frac{3}{2}\big)-\frac{2}{\sqrt{\pi}}\int_0^\infty\bigg[\frac{(1+\epsilon+\frac{\epsilon^2}{2})(\nu+\frac{\nu^2}{2})\epsilon^{\frac{1}{2}}}{\big((\epsilon+\nu)+\frac{(\epsilon+\nu)^2}{2}\big)(\epsilon+\frac{\epsilon^2}{2})}\bigg]\text{d}\epsilon\nonumber\\&\approx&\zeta\big(\frac{3}{2}\big)-2\sqrt{\pi}\nu^{1/2}\bigg[1+\frac{\nu^{1/2}}{2^{3/2}}+{\it{O}}(\nu^{\frac{3}{2}})\bigg].
\end{eqnarray}
Let us use this result at the left hand side of Eqn.(3). In its right hand side, we may have the Taylor expansion about $t=1$ as
\begin{eqnarray}
1/t=1-(t-1)+(t-1)^2+....
\end{eqnarray}
Plugging above two expansions in to Eqn.(3) we get
\begin{eqnarray}
\mu_{\gtrapprox}(t)\approx-kT_c\frac{9\zeta^2(\frac{3}{2})}{16\pi}(\delta t)^2\bigg[1-\big(\frac{1}{4}+\frac{3\zeta(\frac{3}{2})}{8\sqrt{2\pi}}\big)\delta t\bigg]
\end{eqnarray}
up to the second lowest order in $\delta t=t-1$. It should be mentioned that the lowest order term in Eqn.(10) was previously obtained by Bhattacharjee adopting the same trick of Landau and Lifshitz \cite{bhattacharjee,landau}.

Intermediate regime is an overlap of the classical and quantum regimes, and it can be characterized by $t\gnapprox1$. For this case, our approximation is made considering only the first three terms of $\text{Li}_{3/2}(z)$ instead of its all terms in Eqn.(3). Thus solving the cubic equation of $z$ we have
\begin{eqnarray}
\mu_{\gnapprox}(t)\approx t\ln\bigg[-\sqrt{\frac{3}{8}}-\frac{(432\sqrt{3}-162)t^{\frac{3}{2}}}{36\cdot2^{\frac{1}{3}}\cdot3^{\frac{5}{6}}f_1(t)}+\frac{t^{-\frac{3}{2}}f_1(t)}{2\cdot2^{\frac{2}{3}}\cdot3^{\frac{1}{6}}}\bigg],~~
\end{eqnarray}
where $f_1(t)=\big[[36\sqrt{6}-9\sqrt{2}]t^{9/2}+144t^3\zeta(\frac{3}{2})+\sqrt{3}\big([3072\sqrt{3}-864]t^9+[3456\sqrt{6}-864\sqrt{2}]\zeta(\frac{3}{2})t^{15/2}+6912t^6\zeta^2(\frac{3}{2})\big)^{1/2}\big]^{1/3}$. Eqn.(11) eventually is not only applicable for the intermediate regime but also for the classical regime.

For $t\le1$, we have $\mu(t)=0$, and have nothing to be approximated. For $t>1$, we have two independent approximate results in Eqns.(10) and (11). These two results intersect at $t\approx1.711$. Thus for $t>1$, we get the approximate chemical potential as a combination of these two in the following form
\begin{eqnarray}
\mu_{>}(t)\approx\mu_{\gtrapprox}(t)\theta(1.711-t)+\mu_{\gnapprox}(t)\theta(t-1.711),
\end{eqnarray}
where $\theta$ is a unit step function. Now it is time to verify how good is our approximate chemical potential. We can verify it comparing with the (exact) graphical solutions of the chemical potential.

\begin{figure}
\includegraphics{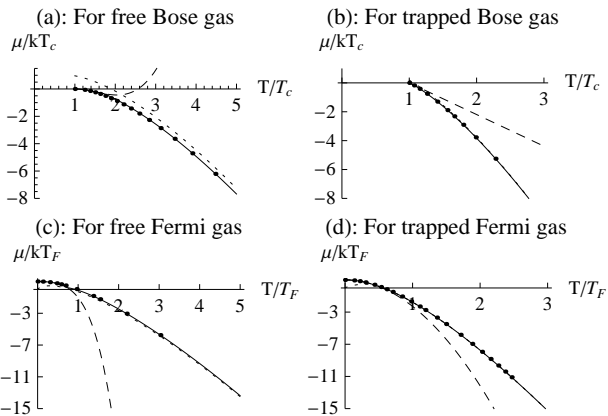}
\caption {Solid lines in FIGs 1a, 1b, 1c \& 1d represent approximate chemical potentials in Eqns.(12), (16), (23) \& (26) respectively. Dotted and dashed lines represent classical and quantum results respectively. Points represent exact graphical solutions of the respective chemical potentials.}
\end{figure}

Let us now outline how to get the graphical solutions. In a figure we plot left hand side of Eqn.(3) with respect to $z$, and right hand side of Eqn.(3) for $t=t_1, t_2, t_3$ etc. Intersecting points of these plots are the solutions ($\{z_i\}$) of $z$ for $t=t_1, t_2, t_3$ etc. Thus we can have a number of graphical solutions $\{\mu_i=kT_i\ln(z_i)\}$ for given values of $t$, and plot them in FIG. 1a. In the same figure we also plot our approximate chemical potential in Eqn.(12), and see that it matches reasonably well with the exact graphical solutions. Hence, we can approximately plot other thermodynamic variables using our approximate chemical potential in Eqn.(12). As an example, we plot specific heat per particle (from Eqn.(2)) in FIG. 2a, and compare with its exact graphical solutions.

That the approximate $c_v$ matches well with the exact graphical solutions gives us courage to adopt our approximation technique to apply for the harmonically trapped ideal Bose gas, free Fermi gas, and for the harmonically trapped ideal Fermi gas as well.

\subsection{For trapped Bose gas}
Chemical potential of the harmonically trapped Bose gas is to be obtained from Eqn.(4). The classical result ($\mu_{\gg}(t)\approx-3kT\text{ln}[t/\zeta^{\frac{1}{3}}(3)]$) for this case can be improved for the intermediate regime considering only the first three terms of $\text{Li}_3(z)$ in Eqn.(4), and it results
\begin{eqnarray}
\mu_{\gnapprox}(t)\approx t\ln\bigg[-\frac{9}{8}-\frac{165t^3}{8f_2(t)}+\frac{3f_2(t)}{8t^3}\bigg],
\end{eqnarray}
where $f_2(t)=\big[261t^{9}+256t^6\zeta(3)+16\sqrt{2}\big(458t^{18}+261t^{15}\zeta(3)+128t^{12}\zeta^2(3)\big)^{1/2}\big]^{1/3}$. On the other hand, in the quantum regime in particular for $t\gtrapprox1$, we should expand $\text{Li}_{3}(z)$ and $1/t$ in Eqn.(4) about $z=1$ and $t=1$ respectively. While the expansion of $\text{Li}_{3}(z)$ can be obtained (using $\text{Li}_3(z)-\zeta(3)=\int_1^z\frac{\text{Li}_2(x)}{x}\text{d}x$ and expanding $\text{Li}_2(x)$ in a manner of obtaining Eqn.(8)) as
\begin{eqnarray}
\text{Li}_{3}(z)=\zeta(3)-\zeta(2)\nu\bigg[1+\frac{\nu}{\zeta(2)}\big(\frac{1}{2}\ln\nu-\frac{3}{4}\big)+...\bigg],
\end{eqnarray}
that of $1/t$ has already been obtained in Eqn.(9). Presence of the logarithmic term in above expansion says that $\text{Li}_3(z)$ is nonanalytic about $z=1$. However, plugging the expansions in Eqns.(14) and (9) in to Eqn.(4) we get \cite{pethick-book}
\begin{eqnarray}
\mu_{\gtrapprox}(t)\approx-3\frac{\zeta(3)}{\zeta(2)}(t-1)
\end{eqnarray}
up to the lowest order in $t-1$. Once again for $t>1$, we have two independent approximate results in Eqns.(15) and (13). These two results intersect at $t\approx1.065$. Combining these two we get our desired result for $t>1$ as
\begin{eqnarray}
\mu_{>}(t)\approx\mu_{\gtrapprox}(t)\theta(1.065-t)+\mu_{\gnapprox}(t)\theta(t-1.065).
\end{eqnarray}
We plot this approximate chemical potential in FIG. 1b, and compare it with its exact graphical solutions.

In the following section we will extend our discussions for the Fermi systems. Although in the classical regime, behavior of a Fermi gas is identical with that of a Bose gas, yet in the quantum regime, they are quite different. This difference needs special care for extending our approximation technique to be applied for the Fermi systems.

\begin{figure}
\includegraphics{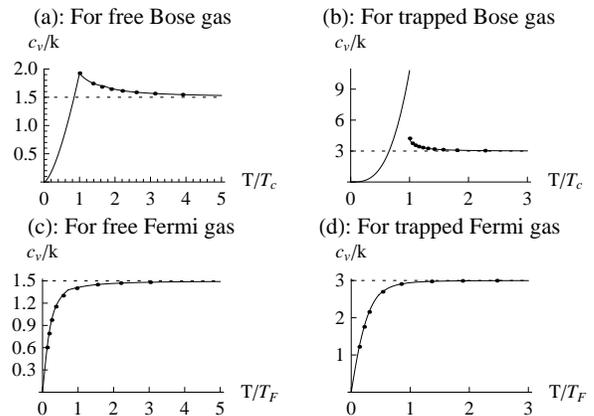}
\caption {Solid lines represent approximate specific heats per particle. Points represent their respective exact graphical solutions. Dotted lines represent classical results.}
\end{figure}

\section{Thermodynamics of Fermi gases}
For an ideal Fermi gas, average number of particles occupying $i$th single particle state is given by the Fermi-Dirac statistics
\begin{eqnarray}
\bar{n}_i=\frac{1}{\text{e}^{(\epsilon_i-\mu)/kT}+1}.
\end{eqnarray}
While for a Bose gas we have defined $t$ as $T/T_c$, henceforth for a Fermi gas, let us redefine $t$ to be $T/T_F$ where $T_F$ is the Fermi temperature. Integrations of right hand side of Eqn.(17) result \cite{pathria}
\begin{eqnarray}
\big(-\text{Li}_{3/2}(-z)\Gamma(5/2)\big)^{2/3}=\frac{1}{t}
\end{eqnarray}
for a free Fermi gas with $T_F=\frac{\hbar^2}{2mk}(6\pi^2\bar{n})^{2/3}$ \cite{pathria}, and
\begin{eqnarray}
\big(-\text{Li}_{3}(-z)\Gamma(4)\big)^{1/3}=\frac{1}{t}
\end{eqnarray}
for a harmonically trapped Fermi gas with $T_F=\frac{\hbar\omega}{k}\big(\Gamma(4)N\big)^{1/3}$ \cite{butts}.

In the following we will present approximate analyses of Eqns.(18) and (19), and will obtain approximate chemical potentials of the two cases indicating suitable suffices for different temperature regimes.

\subsection{For free Fermi gas}
Adopting the previous manner we get the classical result for the free Fermi gas to be $\mu_{\gg}(t)\approx-\frac{3}{2}kT\text{ln}[t\Gamma^{\frac{2}{3}}(5/2)]$, and improved it considering only the first three terms of $-\text{Li}_{3/2}(-z)$ to get the approximate result for the intermediate regime ($t\gtrsim1$) as
\begin{eqnarray}
\mu_{\gtrsim}(t)\approx t\ln\bigg[\sqrt{\frac{3}{8}}-\frac{(432\sqrt{3}-162)\pi^{\frac{1}{2}}t^{\frac{3}{2}}}{36\cdot2^{\frac{1}{3}}\cdot3^{\frac{5}{6}}f_3(t)}+\frac{t^{-\frac{3}{2}}f_3(t)}{2\pi^{\frac{1}{2}}\cdot2^{\frac{2}{3}}\cdot3^{\frac{1}{6}}}\bigg],~
\end{eqnarray}
where $f_3(t)=\big[192\pi t^3-[36\sqrt{6}-9\sqrt{2}]\pi^{\frac{3}{2}}t^{9/2}+\sqrt{3}\big(12288\pi^2t^6+[1152\sqrt{2}-4608\sqrt{6}]\pi^{5/2}t^{15/2}-[864-3072\sqrt{3}]\pi^3t^9\big)^{1/2}\big]^{1/3}$.
On the other hand, for the quantum regime ($t\lnsim1$) as well as for $z\gg1$, we expand $-\text{Li}_{3/2}(-z)$ in Eqn.(18) according to Sommerfeld's asymptotic formula \cite{pathria}
\begin{eqnarray}
-\text{Li}_{j}(-z)&\approx&\frac{[\mu(t)/kT]^{j}}{\Gamma(j+1)}\bigg[1+\sum_{n=1}^\infty\bigg(2\zeta(2n)(1-2^{1-2n})\nonumber\\&&\times\frac{j(j-1)...(j-[2n-1])}{[\mu(t)/kT]^{2n}}\bigg)\bigg].
\end{eqnarray}
Plugging the above expansion for $j=3/2$ in to Eqn.(18) we get \cite{kiess}
\begin{eqnarray}
\mu_{\lnsim}(t)=kT_F\bigg[1-\frac{\pi^2}{12}t^2-\frac{\pi^4}{80}t^4+{\it{O}}(t^6)\bigg].
\end{eqnarray}
For the entire range of temperature, we have two independent approximate results in Eqns.(22) and (20). These two results intersect at $t\approx0.723$. Their combination in the following form
\begin{eqnarray}
\mu(t)\approx\mu_{\lnsim}(t)\theta(0.723-t)+\mu_{\gtrsim}(t)\theta(t-0.723)
\end{eqnarray}
gives us our desired approximate chemical potential. We plot this approximate chemical potential in FIG. 1c, and compare it with its exact graphical solutions.

\subsection{For trapped Fermi gas}
Adopting similar treatment obtaining approximate results of the free Fermi gas, we get the classical and improved classical result for the trapped Fermi gas respectively as $\mu_{\gg}(t)\approx-3kT\text{ln}[t\Gamma^{\frac{1}{3}}(4)]$ and \cite{biswas}
\begin{eqnarray}
\mu_{\gtrsim}(t)\approx kT_Ft\text{ln}\bigg[\frac{9}{8}-\frac{165\times3^{1/3}f_4(t)}{8}+\frac{3^{2/3}}{8f_4(t)}\bigg],
\end{eqnarray}
where $f_4(t)=t/[128-783t^3+16(64-783t^3+8244t^6)^{1/2}]^{1/3}$. For the quantum regime, using the asymptotic formula for $j=3$ we get \cite{butts}
\begin{eqnarray}
\mu_{\lnsim}(t)=kT_F\bigg[1-\frac{\pi^2}{3}t^2+{\it{O}}(t^6)\bigg].
\end{eqnarray}
Once again, for the entire range of temperature, we have two independent approximate results in Eqns.(25) and (24). These two results intersect at $t\approx0.468$. Their combination in the following form \cite{biswas}
\begin{eqnarray}
\mu(t)\approx\mu_{\lnsim}(t)\theta(0.468-t)+\mu_{\gtrsim}(t)\theta(t-0.468)
\end{eqnarray}
gives us our desired approximate chemical potential. We plot this approximate chemical potential in FIG. 1d and compare it with its exact graphical solutions.

\begin{figure}
\includegraphics{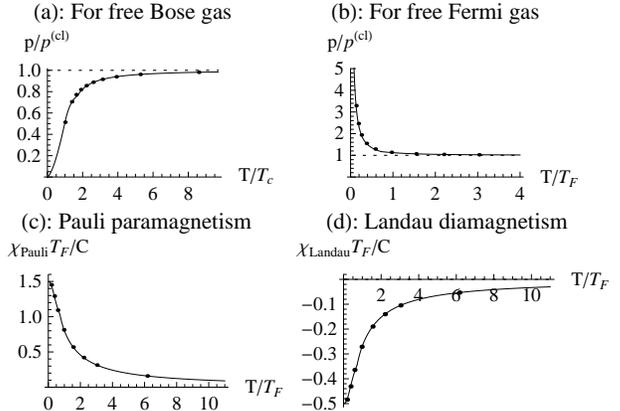}
\caption {Solid lines in FIG 3a and 3b represent equation of state ($p$) in units of $p^{(\text{cl})}=\bar{n}kT$. Solid lines in FIG 3c and 3d represent paramagnetic ($\chi_{\text{Pauli}}$) and diamagnetic ($\chi_{\text{Landau}}$) parts of susceptibilities of the ideal homogeneous Fermi (electron) gas in units of $\text{C}/T_F=\mu_0\bar{n}\mu_B^2/kT_F$. Points represent respective exact graphical solutions.}
\end{figure}

\section{Conclusions}
Although it is impossible to get chemical potentials of the quantum gases in exact closed forms of temperature, yet we have obtained them in approximate closed forms. The closed forms give plausibility to reproduce thermodynamics of the quantum gases for the entire range of temperature. As a claim of plausibility we have plotted specific heats of all the quantum gases in FIG 2, and equations of states (pressures) of the free quantum gases in FIGs 3a \& 3b, and paramagnetic \& diamagnetic parts of the susceptibilities of the ideal homogeneous Fermi gas of spin $1/2$ in FIGs 3c \& 3d using the respective approximate chemical potentials in FIG 1.

We have not shown relationships between specific heats and fugacities for the plots in FIGs 2b, 2c and 2d. These are easily derivable and available in Refs.\cite{pethick-book}, \cite{pathria} and \cite{butts} respectively. We have not also shown the derivations of equations of states and that of the susceptibilities in FIG 3. These are available in standard text books \cite{pathria,bhattacharjee}.

For harmonically trapped systems, we have considered oscillations to be isotropic. Otherwise angular frequency ($\omega$) in our results has be replaced by the geometric mean of $\omega_x$, $\omega_y$, and $\omega_z$ \cite{pitaevskii-book,pitaevskii-rmp1}.

It should have been mentioned that an approximate form of specific heat of the free Bose gas was previously obtained for $T\gtrapprox T_c$ by Wang using Robinson's method which involves Mellin transform as well as analytic continuation of $\text{Li}_j(z)$ \cite{wang,robinson}, and for $T\gnapprox T_c$ by London using virial expansion method \cite{london,pathria}. These methods (or techniques) can also be used for the harmonically trapped Bose gas. On the other hand, approximate form of thermodynamic variables of the free and harmonically trapped Fermi gases can be obtained for $T\lnsim T_F$ using Sommerfeld's asymptotic expansion of $-\text{Li}_j(-z)$, and for $T\gnapprox T_F$ using the virial expansion method \cite{pathria,pitaevskii-rmp2}. While none of the above methods/techniques covers the entire range of temperature, our approximation technique covers the entire range not only for the Bose gases but also for the Fermi gases. However, it would be an interesting problem for undergraduate and graduate students to get a temperature dependent form of specific heat of the trapped Bose gas for $T\gtrapprox T_c$ in the way we have got approximate chemical potentials using the trick of Landau and Lifshitz \cite{landau}. Students can also generalize the virial expansion to get approximate temperature dependence of specific heats of the trapped Bose and Fermi gases.

To conclude, this paper illustrates approximate analyses of polylog functions in demonstrating temperature dependence of thermodynamic quantities of ideal quantum gases for the entire range of temperature in an easy and plausible way. In this paper we have applied our approximation technique only to a few thermodynamic properties of the ideal quantum gases. Our technique can also be applied for other thermodynamic properties (eg. temperature dependence of entropy, energy, etc.) of the ideal and weakly interacting quantum gases \cite{biswas}. All the calculations of this paper have been done within the scope of undergraduate and graduate students. They can extend our approximation technique to get better results including higher order corrections, and generalize our approximation technique to other spatial dimensions.

\section*{Acknowledgment}
This work has been sponsored by the University Grants Commission [UGC] under the D.S. Kothari Postdoctoral Fellowship Scheme {[No.F.4-2/2006(BSR)/13-280/2008(BSR)]}.

\end{document}